\documentclass[12pt]{aastex}
\usepackage{emulateapj5}
\submitted{ApJ}
\newcommand{\mamo}[1]{\mbox{$#1$}}
\newcommand{\unit}[1]{\ifmmode \:\mbox{\rm #1}\else \mbox{#1}\fi}
\newcommand{\sbr}[1]{_{\rm #1}}

\newcommand{\mone}{\mamo{^{-1}}}

\newcommand{\kms}{\unit{km~s\mone}}

\newcommand{\mpc}{\unit{Mpc}}

\newcommand{\hmpc}{\mamo{h_{75}{\mone}\mpc}}
\newcommand{\hmpcinvcub}{\mamo{h_{75}^{3} \mpc^{-3}}}

\slugcomment{\apj, submitted}

\begin{document}

\title{THE OPTICAL LUMINOSITY FUNCTION OF VIRIALIZED SYSTEMS}

\author{Christian MARINONI}
\affil{ Department of Astronomy, Cornell
University, Ithaca, NY}
\affil{Dipartimento di Astronomia, Universit\`{a} degli Studi di
Trieste, Trieste, Italy}
\email{marinoni@astrosun.tn.cornell.edu}

\author{Michael  J. HUDSON}
\affil{Dept. of Physics, University of Waterloo, Waterloo, Canada}
\email{mjhudson@astro.uwaterloo.ca}

\author{Giuliano GIURICIN \altaffilmark{1}}
\affil{ Dipartimento di Astronomia, Universit\`{a} degli Studi di
Trieste, Trieste}
\affil{SISSA, Trieste, Italy}

\altaffiltext{1}{Deceased.}

\begin{abstract}
We determine the optical luminosity function of virialized systems
over the full range of density enhancements, from single galaxies to
clusters of galaxies.  The analysis is based on galaxy systems
identified from the Nearby Optical Galaxy (NOG) sample, which is the
largest, all-sky catalog of 
objectively-identified bound objects presently available.

We find that the $B$-band luminosity function of systems is
insensitive to the choice of the group-finding algorithms and is well
described, over the absolute-magnitude range $-24.5 \leq M + 5 \log
h_{75} \leq -18.5$, by a Schechter function with
$\alpha\sbr{s}=-1.4\pm 0.03$, $M\sbr{s}^{*} - 5 \log h_{75} =-23.1 \pm
0.06$ and $\phi\sbr{s}^{*}=4.8 \times 10^{-4}\;\hmpcinvcub$ or by a
double power law: $\phi_{\rm pl}(L\sbr{s}) \propto L\sbr{s}^{-1.45 \pm
0.07}$ for $L\sbr{s}< L\sbr{pl}$ and $\phi(L\sbr{s}) \propto
L\sbr{s}^{-2.35 \pm 0.15}$ for $L\sbr{s} > L\sbr{pl}$ with
$L\sbr{pl}=8.5 \times 10^{10} h_{75}^{-2} L_{\odot}$, corresponding to
$M\sbr{s}- 5 \log h_{75} = -21.85$.  The characteristic luminosity of
virialized systems, $L\sbr{pl}$, is $\sim 3$ times brighter than that
($L^{*}\sbr{gal}$) of the luminosity function of NOG galaxies.  Our
results show that half of the luminosity of the universe is generated
in systems with $L\sbr{s}< 2.9 L^{*}\sbr{gal}$ and that $10$\% of the
overall luminosity density is supplied by systems with $L\sbr{s}>30
L^{*}\sbr{gal}$.

We find a significant environmental dependence in the luminosity
function of systems, in the sense that overdense regions, as measured
on scales of $5 h^{-1} \mpc$, preferentially host brighter, and
presumably more massive, virialized systems.

\end{abstract}


\section{INTRODUCTION}

From a theoretical perspective, dark matter halos are the basic
building blocks of the structure in the Universe.  Observationally,
halos must be identified with virialized galaxy ``systems'', ranging
from single dwarf galaxies to rich clusters of galaxies.  The missing
link between the two is the mass-to-light ratio, which in general is a
function of mass or luminosity.  As a first step toward understanding
this relationship, in this paper we determine the luminosity function,
not of individual galaxies, but of entire virialized systems.  This
can be compared with the luminosity functions theoretically predicted
by semi-analytic models of galaxy formation and with other halo
statistics (Marinoni \& Hudson 2001), such as the mass function.

While the luminosity function (LF) of cluster and group galaxies has
been extensively studied in literature (Iovino et al.\ 1993; Marzke \&
da Costa 1997; Lumsden et al.\ 1997; Gaidos 1997; Rauzy, Adami, \&
Mazure 1998; Bromley et al.\ 1998; Ramella et al.\ 1999; Marinoni et
al. 1999), only few detailed studies have examined the LF of galaxy
systems. Gott \& Turner (1977, GT) used data for 103 groups identified
in two dimensions (i.e. from the angular coordinates of galaxies only)
to determine the functional shape of the observed group LF.  Bahcall
(1979), combining the GT data with the Abell (1958) clusters,
attempted to extrapolate the observed luminosity distribution into the
rich-cluster domain.  These early studies, however, suffered from a
number of weaknesses: groups identified only in projection on the sky,
small-number statistics, and uncertain completeness and corrections.
These weaknesses were largely overcome by MFW, who determined the LF
of systems using 163 groups (with $\geq 3$ members) identified in the
CfA1 redshift survey (Davis et al.\ 1982; Huchra et al.\ 1983), the
best data set available at that time. The quality and quantity of the
data have improved considerably since those studies, allowing a more
robust determination of the LF of
virialized systems (hereafter VSLF). 

Our VSLF is based on the  Nearby Optical Galaxy (NOG) catalog
(Marinoni 2001), a nearly all-sky, complete, magnitude-limited sample
of $\sim$ 7000 bright and nearby galaxies.  In order to assess the
effects of grouping algorithms, we will analyze three catalogs of
groups, all based on NOG but extracted using different algorithms and
selection criteria (Giuricin et al.\ 2000).  When grouped, NOG
contains $\sim$2800 isolated galaxies and $\sim$ 1100 systems
($\sim$600 binaries and $\sim$500 groups with at least three members).
This group catalog combines a representative volume ($cz < 6000 \kms$,
nearly all-sky) of the Universe with a high comoving density of
galaxies to allow excellent statistics in the group regime.

This paper is the fifth in a series (Marinoni et al.\ 1998, Paper I;
Marinoni et al.\ 1999, Paper II; Giuricin et al.\ 2000, Paper III;
Giuricin et al.\ 2001, Paper IV;) in which we investigate the
properties of the large-scale structures as traced by the NOG
sample. The outline of our paper is as follows: in \S 2 we review and
summarize the identification procedures of NOG galaxy systems and
discuss some specific properties of NOG groups.  In \S 3 we describe
our method for calculating the VSLF and apply it to our sample.
Results are summarized in \S 4.

Throughout this paper, the Hubble constant is taken to be 75 $h_{75}$
km s$^{-1}$ Mpc$^{-1}$ and the recession velocities $cz\sbr{LG}$ are
evaluated in the Local Group rest frame.

\section{The NOG groups: identification and spatial distribution}

The NOG sample is a statistically controlled, distance-limited
($cz\sbr{LG}\leq 6000 \kms$) and magnitude-limited ($B \leq 14$)
complete sample of more than $7000$ optical galaxies.  The sample
covers 2/3 (8.27 sr) of the sky ($|b|>20^{\circ}$), a volume of $1.41
\times 10^{6}\hmpcinvcub$ and has a redshift completeness of 98\% (see
Marinoni 2001).  In this paper we use the small extension of the NOG
(for a total of 7232 galaxies) containing 156 additional galaxies
which have rough estimates of magnitudes (see Paper III for details).
Note that, in contrast to most previous studies of large-scale
structure, the magnitudes used in this paper are homogenized total
blue magnitudes, given in the standard system of the RC3 catalog (de
Vaucouleurs et al.  1991) and fully corrected for Galactic and
internal extinction and K-dimming.

Redshift maps present two principal distortions: i) small-scale
perturbations caused by random velocities in clusters of galaxies
which produce the so-called ``Fingers of God'' i.e.\ a radial
stretching in galaxy maps pointed at the observer and ii) large-scale
perturbations caused by large overdensities leading to coherent {\em
bulk} motions.  For the purposes of this paper, the latter effect of
large-scale peculiar velocities are not severe; these can be easily
treated via the flow models derived in Paper I.

We corrected for the Finger-of-God effect in Paper III, where galaxy
systems were identified by means of different objective, group-finding
algorithms i.e.\ the widely-used hierarchical (H) and percolation (P)
(or {\em friends-of-friends\/}) algorithms. The P algorithm (Huchra \&
Geller 1982) identifies as members of galaxy aggregations the galaxies
which have their transverse separation $D_{ij} \leq D_0 \cdot R$ and
line-of-sight velocity difference $cz_{ij} \leq cz_0 \cdot R$, where R
is a scaling parameter which takes into account the decrease of the
magnitude range of the luminosity function sampled at increasing
distance.  In Paper III, two variants of the percolation method were
used to identify groups.  In one variant (hereafter denoted as
"scaled" friends-of-friends or P2, following Paper III), typical
values for the linking parameters, at the median depth of the sample
($ \sim 4000$ \kms), are 434 \kms\ in velocity and 0.89 $\hmpc$ in
transverse separation, and both link parameters were then suitably
scaled with distance to give a density threshold of $\frac{\delta
n}{n}=80$.  In the other variant (hereafter denoted simply as
percolation or P1), only the transverse separation link parameter was
scaled with distance, while the velocity link parameter was fixed at
the value of 350 \kms.  Given the limited range of redshift
encompassed by the NOG, this choice was used to approximate a slow
scaling of the velocity link parameter with distance, as suggested by
cosmological N-body simulations (e.g., Nolthenius \& White 1987).

The hierarchical clustering method introduced by Materne (1978) and
revised by Tully (1980) is an algorithm which generates a hierarchical
sequence of systems organized by some {\em affinity} parameter.  In
paper III, the authors adopted $\rho_{L}=8\times 10^{9} L_{\odot}$
Mpc$^{-3}$ (corresponding to a luminosity density contrast
$\frac{\delta \rho_L}{\rho_L}=45$) as the limiting luminosity density
parameter used for cutting the hierarchy and defining groups.

For both algorithms, the adopted values of the group-selection
parameters were chosen after a search in the parameter space guided by
numerical simulations in order to obtain realistic and homogeneous
catalogs of groups.  Since all three catalogs are based on the same
galaxy sample, this allows us to investigate systematic effects
related to the choice of grouping algorithm.

Most of the NOG galaxies ($\sim$60\%) are found to be members of
galaxy binaries (which comprise $\sim$15\% of galaxies) or groups with
at least three members ($\sim$45\% of galaxies).  About 40\% of the
galaxies are left ungrouped (isolated galaxies).

In Paper III, the similarity between the catalogs of groups was
demonstrated using the fraction of members in common to groups of
different catalogs.  In this paper, we also show that the three
catalogs of groups show similar statistical distributions and trace
similar large-scale structures.  Figure \ref{figa} shows the
distributions of P2 groups on the sky.  The smoothed group number
density contrast at a given point {\bf r} (in units of velocity) has
been computed by summing the contribution from all the $i$ groups of
our catalog:

\begin{equation} \frac{\delta n}{n}({\bf r}, R\sbr{s})=\frac{1}{n}\sum_{i}
\frac{W\Big(\frac{|{\bf r}-{\bf r}_i|}{R\sbr{s}}\Big)}{S(r_{i})}-1
\end{equation}

where \begin{equation} n=\frac{1}{V}\sum_{i} S({\bf r}_{i})^{-1}
\end{equation}

\begin{equation} W\Big(\frac{|{\bf r}|}{R\sbr{s}}\Big)=\frac{1}{(2 \pi
R\sbr{s}^2)^{3/2}}e^{-\frac{r^2}{2R\sbr{s}^2}} \label{gauss} \end{equation}

\begin{equation} S(r)=\frac{\int_{-\infty}^{m\sbr{lim}-5 \log r -15 }
\phi(M)dM}{\int_{-\infty}^{\infty} \phi(M)dM \label{sel}}
\end{equation} are the mean number density, smoothing and selection
functions  respectively,  and where     $m\sbr{lim}=12$ is the   limiting
apparent magnitude of completeness of  the group catalog (see \S 3.2).
Smoothing  is performed in three dimensions  with  a variable Gaussian
smoothing length  $R\sbr{s}$ given   by  the mean  inter-group   separation
ranging from 300 to 670 \kms.

\vbox{%
\begin{center}
\leavevmode
\hbox{%
\epsfxsize=18.9cm
\epsffile{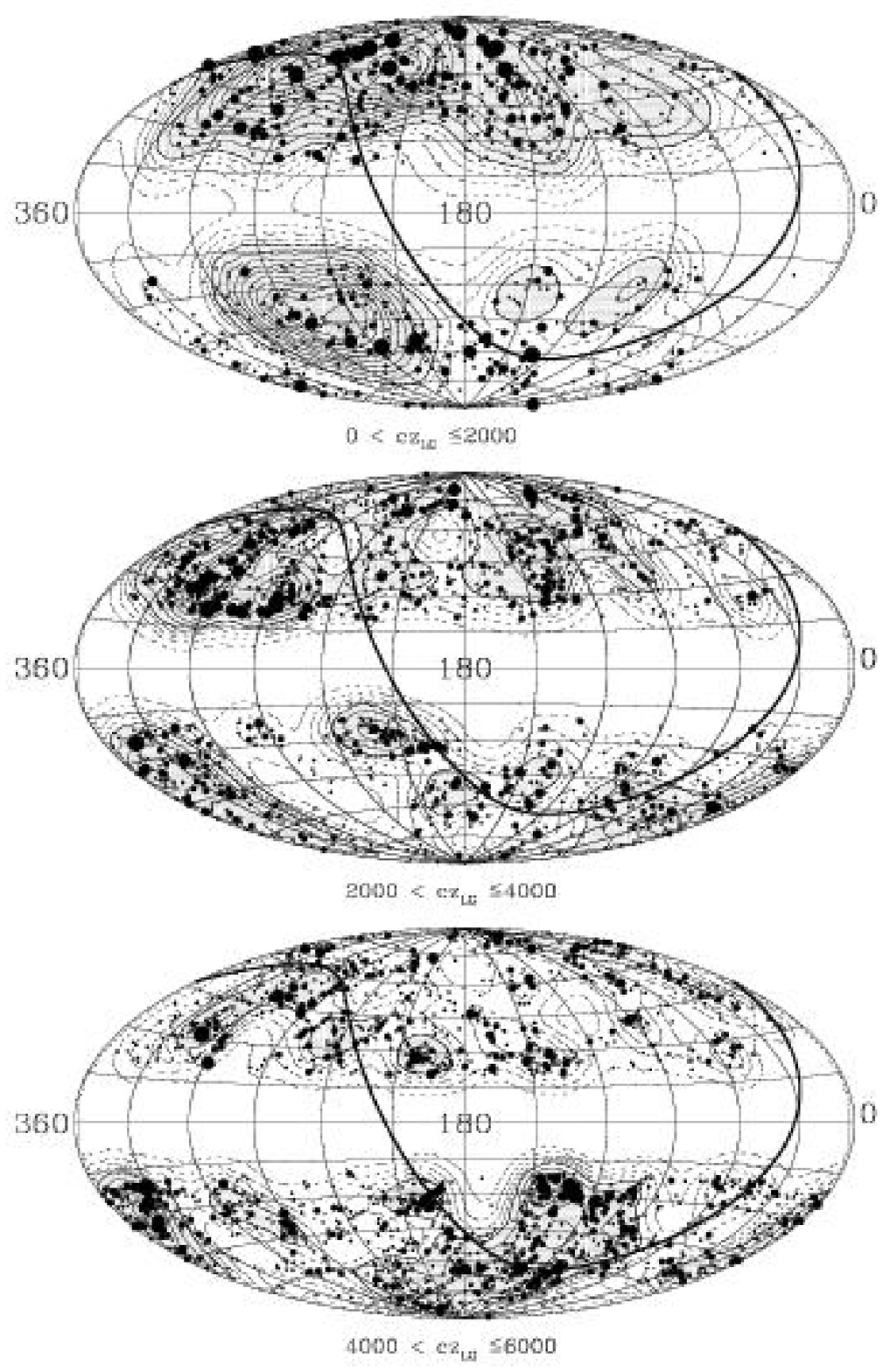}}
\begin{small}
\figcaption{%
NOG galaxy systems reconstructed using the
scaled {\it friends of friends} (P2) algorithm are shown in equal-area
Aitoff projection on the sky using Galactic coordinates.  The region
devoid of galaxies is the zone of avoidance ($|b|<20^{\circ}$).  The
thick line is drawn at the celestial equator, $\delta=0^{\circ}$.
Galaxy systems are indicated by filled circles, where the symbol size
scales with the number of members of each groups. Also shown is
the smoothed distribution of systems on shells at redshift 1000 \kms
(upper), 3000 \kms\ (center), 5000 \kms\ (lower).  Smoothing is
performed in 3D with a variable Gaussian smoothing length given by the
mean group separation (which is given by 309, 480 and 622 \kms
respectively).  The heavy contour denotes the mean density.  Dashed
contours represent densities below the mean (spaced with $\Delta
(\frac{\delta n}{n})=0.2$ intervals).  Regions in excess of the mean
are in gray scale with contours spaced with $\Delta (\frac{\delta
n}{n})=0.4$ intervals.
\label{figa}}
\end{small}
\end{center}}

\vbox{%
\begin{center}
\leavevmode
\hbox{%
\epsfxsize=18.9cm
\epsffile{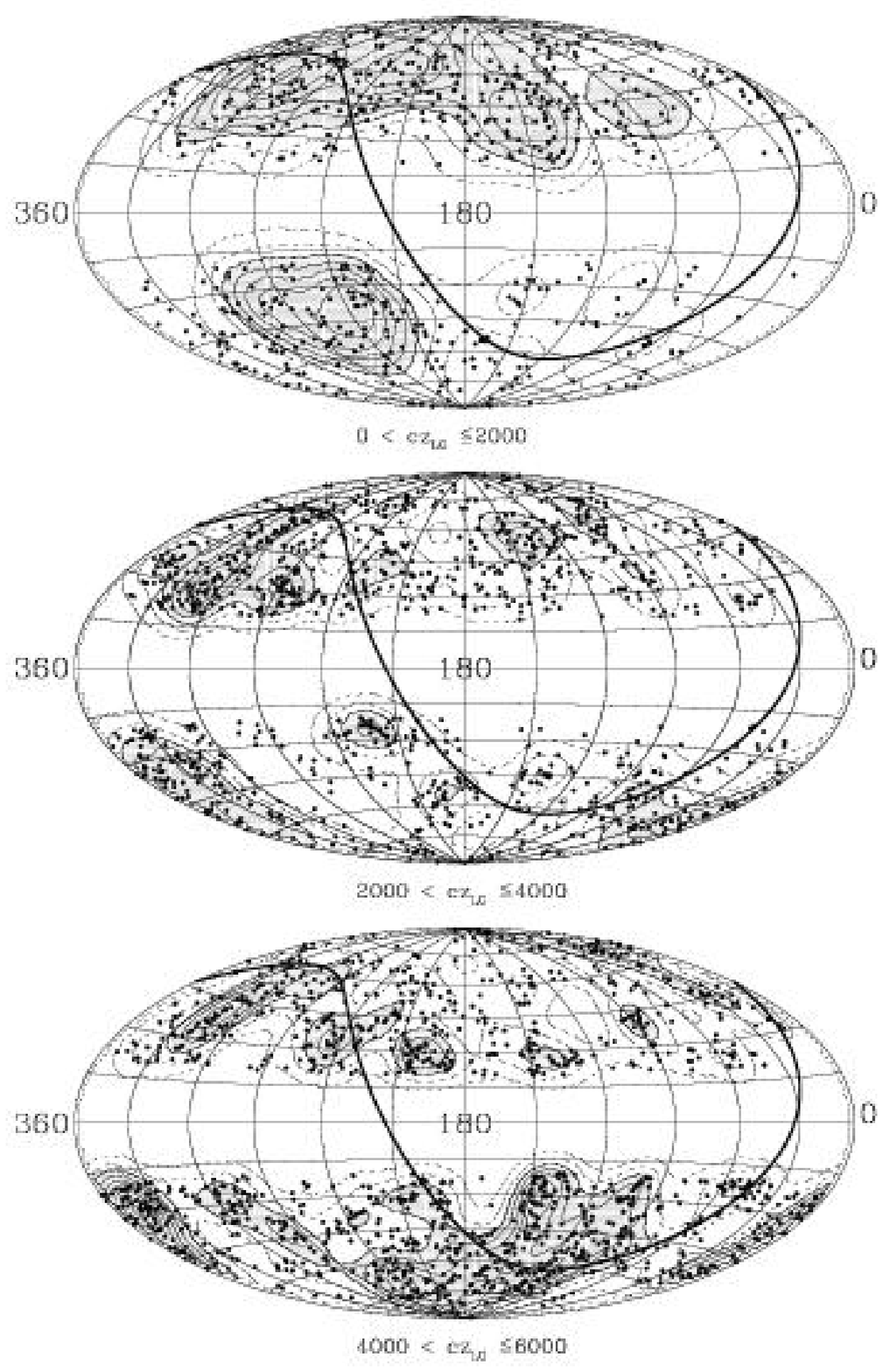}}
\begin{small}
\figcaption{%
NOG isolated galaxies, i.e. galaxies left ungrouped after running the
scaled friends of friends algorithm (P2), are shown in equal-area
Aitoff projection on the sky using Galactic coordinates.  The region
devoid of galaxies is the zone of avoidance ($|b|<20^{\circ}$).  The
thick line is drawn at the celestial equator, $\delta=0^{\circ}$.
Also shown is the smoothed distribution of systems on shells at
redshift 1000 $\kms$ ({\em upper}), 3000 $\kms$ ({\em center}), 5000
$\kms$ ({\em lower}).  Smoothing is performed in 3D with a variable
Gaussian smoothing length given by the mean group separation (which is
given by 309, 480 and 622 $\kms$ respectively).  The heavy contour
denotes the mean density.  Dashed and solid contours represent
densities below and above the mean and are spaced with
$\Delta(\frac{\delta n}{n})=1/3$ intervals.  Regions in excess of the
mean are in gray scale.
\label{figaa}}
\end{small}
\end{center}}
\clearpage

\vbox{%
\begin{center}
\leavevmode
\hbox{%
\epsfxsize=19.9cm
\epsffile{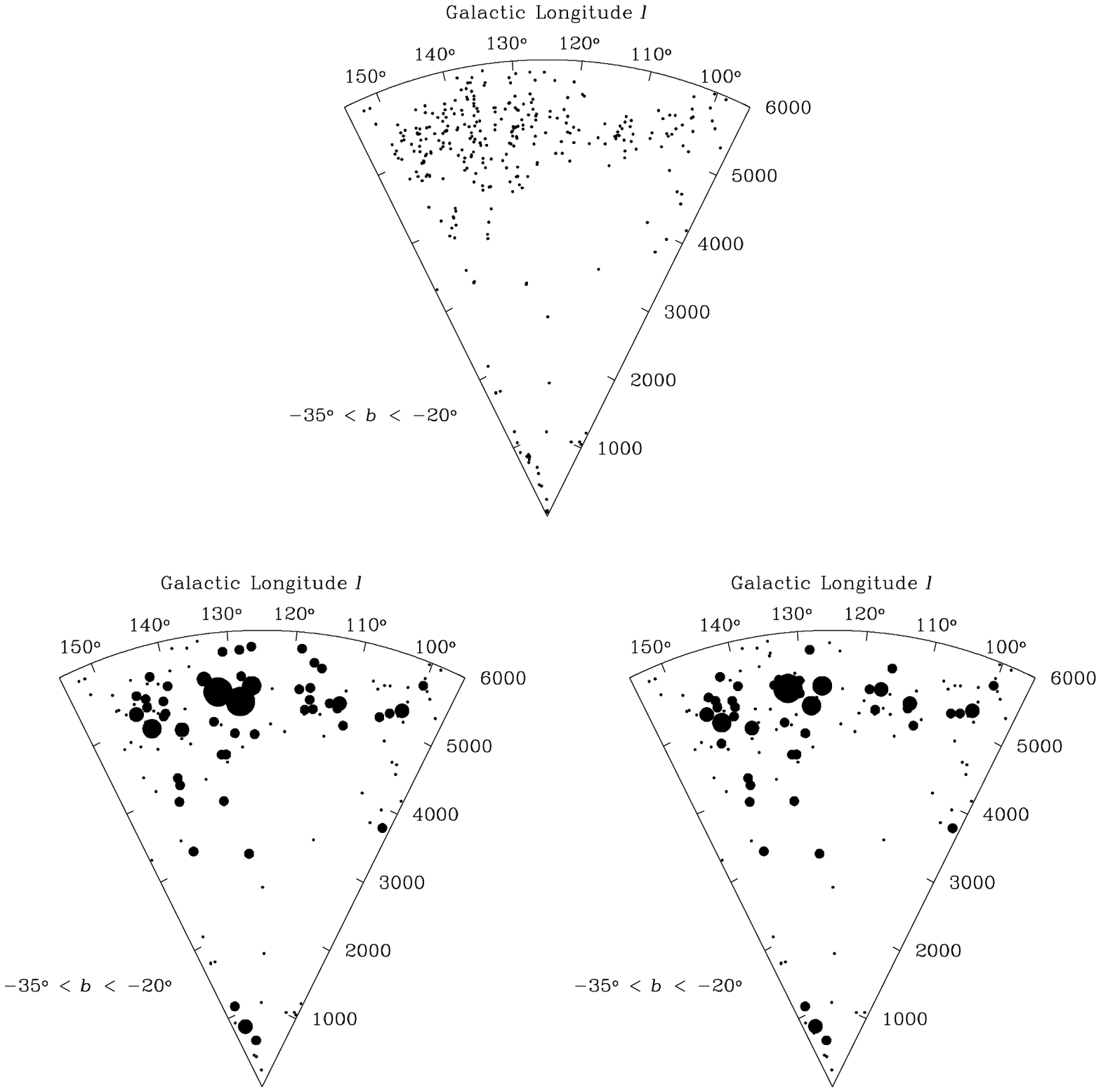}}
\begin{small}
\figcaption{%
Cone diagrams of the distribution of NOG galaxies in the
Perseus-Pisces region: the distance coordinate is the recession
velocity (in the Local Group frame), angular coordinates is Galactic
longitude (l).  {\em Upper:}) distribution of galaxies before
grouping; {\em Lower Left:}) distribution of objects after the
application of the hierarchical (H) group-finding method; {\em Lower
Right:}) distribution of objects after the application of the {\it
friends of friends} (P1) group-finding algorithm. The same dot scaling
as in Figure 1 is used to represent groups with different number of
members.
\label{figc}}
\end{small}
\end{center}}
\clearpage

\vbox{%
\begin{center}
\leavevmode
\hbox{%
\epsfxsize=19.9cm
\epsffile{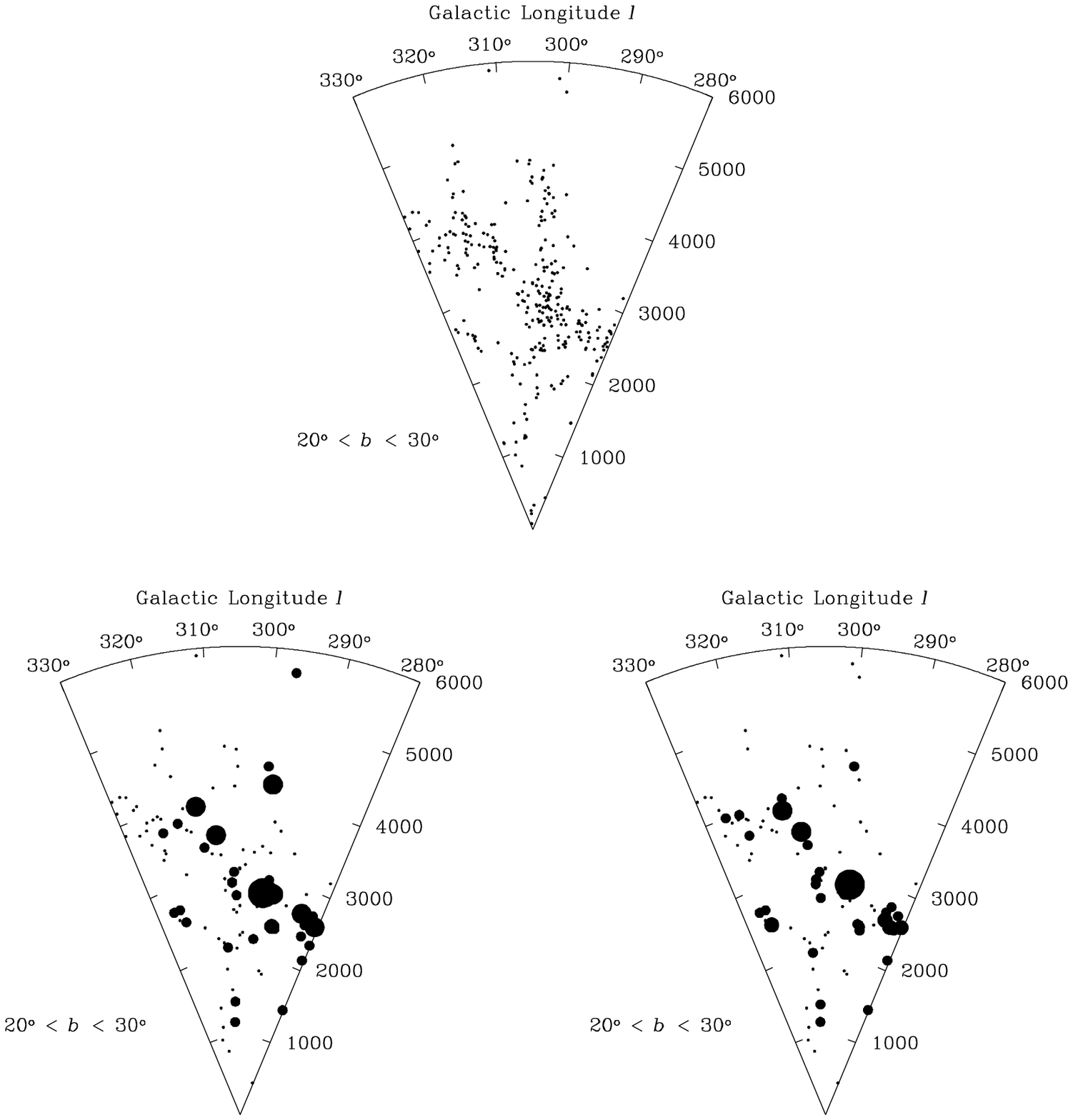}}
\begin{small}
\figcaption{%
As in Figure 2 but for the Hydra-Centaurus region.
\label{figd}}
\end{small}
\end{center}}
\clearpage

A comparison between Fig.\ 1 and Fig.\ 2 shows that groups delineate
the same large-scale structures as do galaxies but with a different
density contrast, i.e. groups are ``biased'' with respect to galaxies.
In a separate paper (paper IV), we presented a detailed analysis of
the clustering of NOG groups.

In Figures \ref{figc} and \ref{figd}, we use cone diagrams in order to
illustrate graphically the performance of two of our reconstruction
algorithms (the H and P1 methods) in two interesting regions of the
sky, the Perseus-Pisces and Hydra-Centaurus superclusters.  These
plots show that the different catalogs of groups have consistent
spatial distribution.  The top cone shows the ungrouped distribution
of galaxies in redshift-space.  Virialized systems show up as
``Fingers of God'' which are elongated along the observer's line of
sight.  Most groups as well as isolated galaxies are recovered by both
methods.  Moreover, these plots demonstrate how well our optimized
grouping algorithms perform across a wide range of density
enhancements and across the different large-scale structures in which
the galaxies are embedded (i.e.  orthogonal and parallel to the
observer's line of sight).

\vbox{%
\begin{center}
\leavevmode
\hbox{%
\epsfxsize=8.9cm
\epsffile{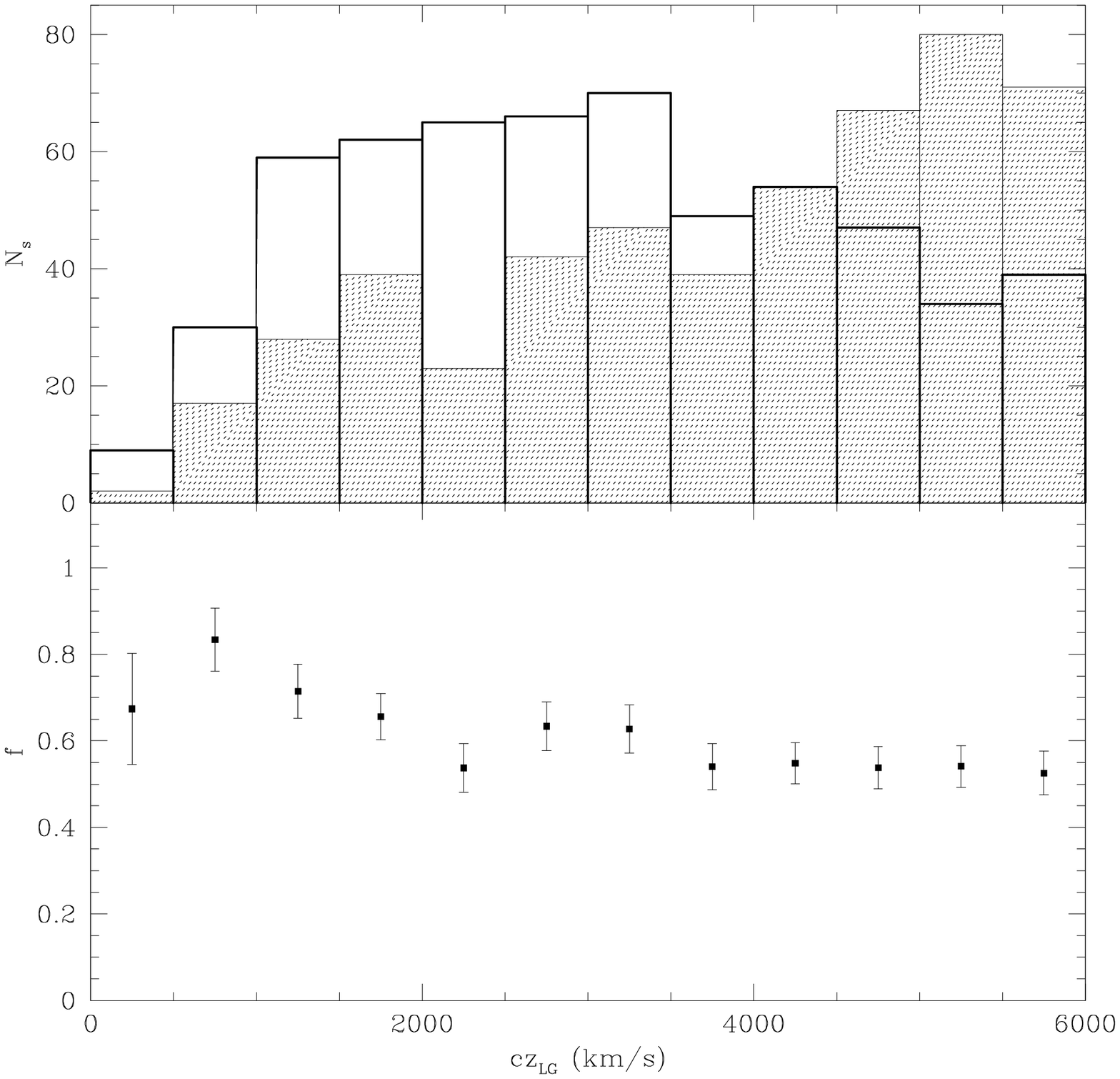}}
\begin{small}
\figcaption{%
{\em Upper:} distribution of 1093 hierarchical binaries and groups as
a function of distance in the northern and southern (shaded
histograms) Galactic hemisphere.  {\em Lower:} fraction of galaxies in
hierarchical groups as a function of distance across the overall NOG
volume. Data are binned in 500 \kms\ distance intervals.
\label{fige}}
\end{small}
\end{center}}

Figure \ref{fige} illustrates that the fraction of NOG galaxies that
are grouped is roughly a constant, independent of distance, although,
due to large-scale clustering, the number of systems is an increasing
function of distance in the southern galactic hemisphere and a
decreasing one in the north.  This indicates that, although we lose an
increasing number of galaxies and groups with distance, due to the
magnitude-limited sample, our group reconstruction methods do not
suffer from any serious distance-dependent bias.  Fig.  \ref{fige}
shows the results for H groups, but the other catalogs of groups give
similar results.

\section{The Luminosity Function of Virialized Systems}

The NOG groups are drawn from a magnitude-limited galaxy sample.  In
order to calculate the VSLF, we regard all systems as aggregates of
galaxies much in the same way as galaxies are aggregates of stars.
However, the crucial difference is that while galaxy fluxes can be
directly measured with some device down to a limiting surface
brightness, the flux of a system has to be inferred from members above
a given apparent magnitude.  When calculating the LF of galaxies, we
must first correct for missing stars, which fall below the surface
brightness limit, and then correct for the missing galaxies which fall
below the magnitude limit of the catalog.  In the same way, when
constructing the LF of systems, we must first correct for missing
galaxies (which fall below the magnitude limit of the survey) and then
correct for missing systems.

We assume here that every system contains galaxies drawn from a
universal galaxy LF, and thus has a large number of faint members.
Consequently, even a single galaxy should be considered as the sole
member observed above the apparent magnitude limit of a group, and
thus requires a correction for unseen luminosity.  By applying a
luminosity correction to all systems, we obtain a ``virialized
systems'' catalog.  We stress that the result is fundamentally
different from the group catalog from which it is derived.  The group
catalog suffers from the well-recognized problem that two groups with
N visible elements identified (from a magnitude-limited galaxy sample)
at two different distances correspond to potential wells of different
masses.  The systems catalog is more physical in the sense that it
corresponds to halos with total absolute luminosity L.

To overcome these difficulties and estimate the luminosity function of
systems, we must 1) assume a model describing how the missed galaxies
are clustered around ``seeds'' defined by NOG systems, 2) check the
photometric completeness of the resulting catalog in order to ensure
that its luminosity distribution is representative of the group
population and not only of our specific sample; 3) use a statistical
estimator which is independent of density fluctuations because we know
that groups tend to cluster at least as strongly as do galaxies (see
Fig. \ref{figa}).

\subsection{The Luminosity Selection Function}

We begin by introducing the statistical method used for estimation of
the total luminosity $L\sbr{s}$ of 
systems.

The NOG galaxy LF, corrected for Malmquist bias and calculated after
correcting for clustering is described by a Schechter-type function (Schechter
1976) with the following shape parameters: $\alpha\sbr{gal} = -1.10
\pm 0.03 $, $M\sbr{gal}^{*}-5 \log h_{75}=-20.61 \pm 0.04$ (Marinoni
2001).

Marinoni (2001) also showed that the LF of galaxies {\em in\/} groups
is marginally consistent with the field galaxy LF.  Therefore, we
assume that the observed luminosity in each bound system is a random
realization of that portion of the universal LF $\phi(L)$ which lies
above the local magnitude limit at the group distance, evaluated from
the median redshift of its galaxy members.  Specifically,
$L\sbr{lim}=10^{0.4 (M_{\odot}-M\sbr{lim})}L_{\odot}$, where
$M_{\odot} = + 5.48$ mag and $M\sbr{lim}= m\sbr{lim}-5 \log \langle
cz_{lg}\rangle -15 + 5 \log h_{75}$.

\vbox{%
\begin{center}
\leavevmode
\hbox{%
\epsfxsize=8.9cm
\epsffile{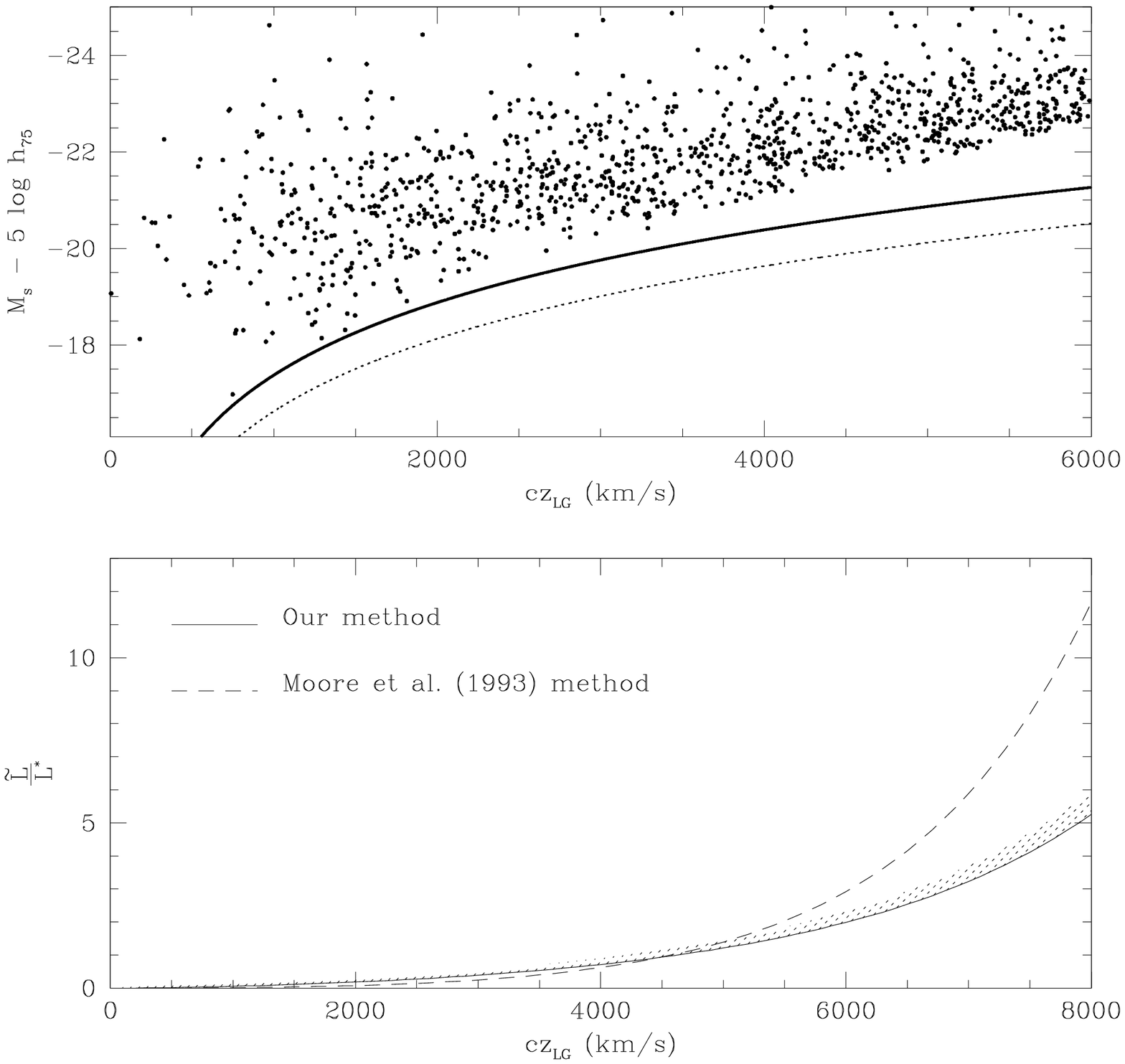}}
\begin{small}
\figcaption{%
{\em Upper:} Spaenhauer diagram (Spaenhauer 1978) showing magnitude
versus redshift for groups with $N \geq 2$ members.  The absolute
magnitude of the systems is the total one, i.e.  corrected for unseen
members.  The solid line is the expected prediction for groups without
any correction applied and corresponds to the following apparent
variation in absolute magnitude $M_s=m^{lim}\sbr{s}-5 \log \langle
cz\sbr{LG} \rangle -15 + 5 \log
h_{75}\;\;\;\;(m^{lim}\sbr{s}=13.25)$.  The dotted line is the
visibility function of the NOG galaxy catalog and corresponds to an
apparent magnitude cutoff $m^{lim}\sbr{gal}=14.$ {\em Lower:} The
additive luminosity corrections per galaxy, in units of $L^{*}\sbr{gal}$ as given
by our luminosity selection function is compared with the MFW method,
using the magnitude system adopted in the NOG and the magnitude
cut-off $m^{lim}\sbr{gal}=14$. The hatched area shows how our luminosity
selection function varies when it is computed using a power-law
approximation at the faint end ($M$ fainter than $-16$) with an
exponent in the range from $-1.1$ to $-1.8$.
\label{figf}}
\end{small}
\end{center}}

In order to recover the total number of galaxies in a system, the
observed number density of galaxies is weighted  by $w_{N}(r)=1/S(r)$,
where $S(r)$ is the selection function of a magnitude-limited sample
(cf.\ eq.\ \ref{sel}), yielding a corrected number density
$n(r)=n\sbr{obs}(r) w_{N}(r)$.  For luminosity, we can write the total
expected luminosity of a system as

\begin{equation} L\sbr{s}(r) = L\sbr{obs}(r)+ \tilde{L}(r) = L\sbr{obs}(r)
w_{L}(r) \label{mco} \end{equation} 
where $\tilde{L}$ denotes the unseen luminosity from galaxies below
the magnitude limit , and $w_{L}(r)$ is the {\em luminosity-density
weighting function} (see also Gourgoulhon, Chamaraux, \& Fouqu\'e
1992)

\begin{equation} w_{L}=\frac{\int_{0}^{\infty} L \phi(L) dL}
{\int_{L\sbr{lim}(r)}^{\infty} L \phi(L)dL}.  \label{lsf} \end{equation}

As $r$ goes to zero, the luminosity weighting function $w_{L}(r)$
reduces to 1 as desired.  With this weighting scheme, we obtain an
estimate of the absolute luminosity of groups which is independent of
the magnitude limit of the catalog in which systems are identified, is
independent of density fluctuations (i.e.  independent of the LF
normalization factor $\phi^*$) and is less sensitive to the faint-end
uncertainties and environmental dependencies of the galaxy LF.

There are other ways in which the total group light can be recovered.
For example, MFW assumed the observed number of galaxies in each group
to be a random sample of the observed portion of the LF and used an
additive correction term of the form
$\tilde{L}=N\sbr{obs}\int_{0}^{L\sbr{lim}(r)} L\phi(L)dL
/(\int_{L\sbr{lim}(r)}^{\infty} \phi(L)dL)$.

Our correction and that of MFW are compared in Figure \ref{figf} where
we plot the fractional expected correction in $L\sbr{gal}^{*}$ units
as a function of distance for a single galaxy with luminosity
$L\sbr{gal}^*$.  Note that we do not expect these two corrections to
be identical, since our correction is applied to the observed
luminosity, whereas that of MFW is applied to the observed number.

We have tested our correction by moving nearby groups to progressively
greater distances.  However, at large distances our approach has the
advantage that it depends very weakly on the adopted Schechter
parameters because, at the faint end, the correction does not weight
using the number density of dwarf galaxies, but using their luminosity
density.  (Note that our correction does not diverge using 0 as the
lower limit of integration.)  Thus, even if the galaxy function
suffers some environmental dependence, i.e.  becoming steeper at the
faint end in clusters (Marzke \& da Costa 1997), this correction is
insensitive to this change.  This can be seen explicitly in Figure
\ref{figf}, where we show the effect on our correction when we modify
the faint-end behavior of the Schechter galaxy luminosity function by
adding a power-law term with an exponent equal to $-1.8$ for
luminosities fainter than $M = -16$ (see Zucca et al.\ 1997).  The
correction to the total system luminosity is small ($< 20\%$) and is
not a strong function of distance. As noted above, we find a shallow
slope ($-1.1$) for NOG galaxies and find no evidence for a large
difference in the NOG B-band LF parameters for field galaxies versus
galaxies in groups and clusters.  Thus we can take the above
correction as a conservative upper limit on such systematic effects.

\subsection{Sample Completeness}

We have applied eq. \ref{lsf} to all systems, including isolated
galaxies.  This yields three ``virialized system'' (VS) catalogs of
objects which, depending on the catalog of groups used, contain $\sim
4000$ objects ($\sim$ 2800 systems with one observed member and
$\sim$1100 groups with at least 2 members).  For example, the
hierarchical VS sample is composed by 2828 isolated galaxies and 1093
groups (with 4404 members) for a total of 3921 ``systems''.  There are
691 binary systems, 297 systems with $3 \leq N <5$ members and 195
systems with $N \geq 5$ members.

The large number of VS considered is not by itself a guarantee that
our catalogs are photometrically complete.  Instead, it is necessary
to analyze the scaling of the number of systems as a function of the
reconstructed magnitude and consider as suitable systems for the LF
estimate only those for which the observed logarithmic scaling follows
roughly the expected Euclidean $N\sbr{s} \propto 10^{0.6 m\sbr{s}}$
behavior (no cosmological corrections are needed within the NOG
volume).  Each VS subsample (singles, binaries, etc.) has been
analyzed with this method.  As an example, in Figure \ref{figg}, we
plot integral and differential (independent bins) logarithmic counts
for the sample of 195 hierarchical systems containing $N\geq 5$
members.  

\vbox{%
\begin{center}
\leavevmode
\hbox{%
\epsfxsize=8.9cm
\epsffile{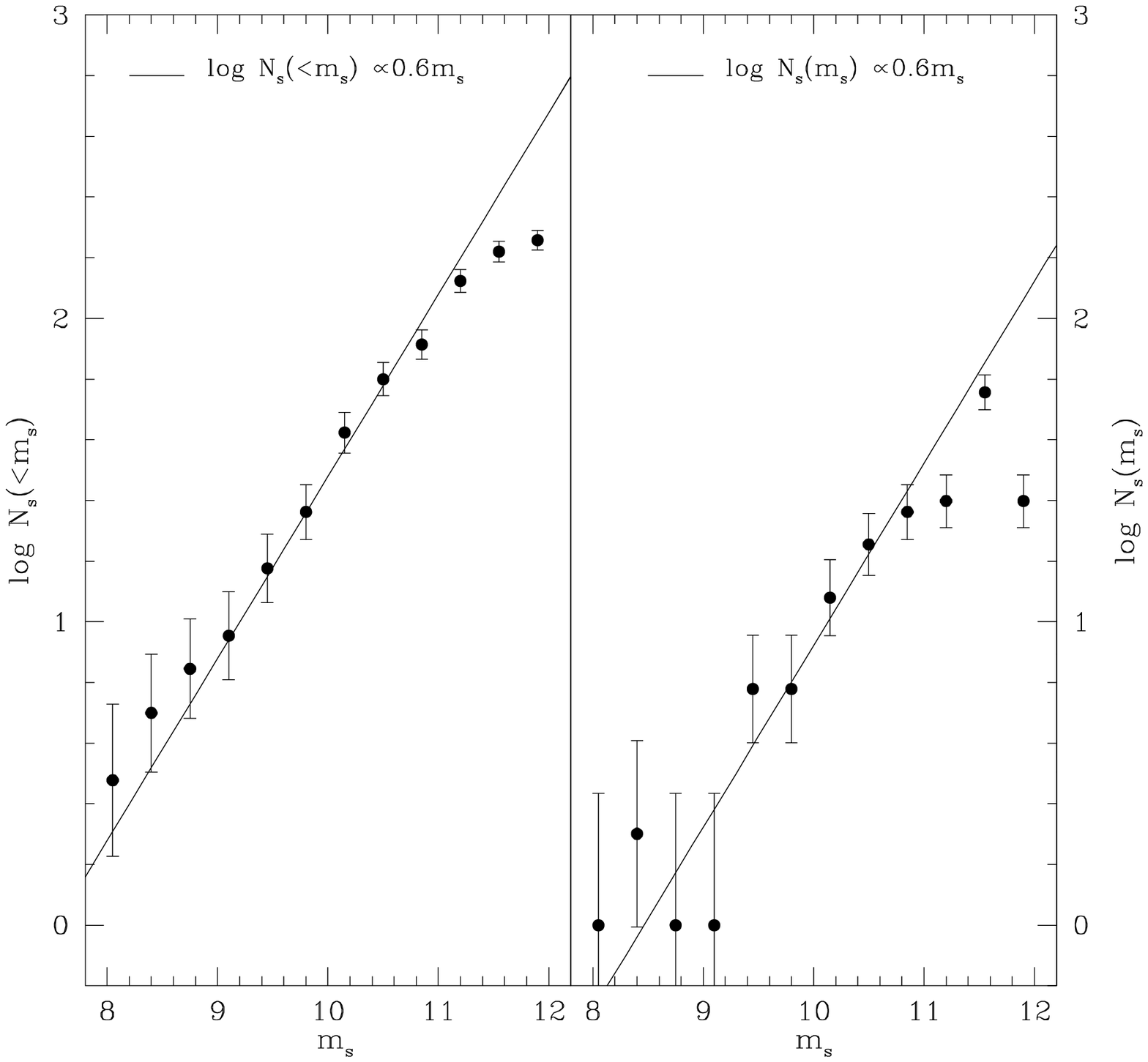}}
\begin{small}
\figcaption{%
Integral ({\em left}) and differential ({\em right}) counts (binned in
0.4 magnitude intervals) for NOG groups with $N\sbr{gal}\geq 5$
members. Apparent magnitudes have been corrected with the luminosity
selection function.  Bars represent $\pm 1 \sigma$ Poisson
errors. Solid lines indicate the expected prediction in a volume with
an Euclidean geometry and a homogeneous distribution of systems.
\label{figg}}
\end{small}
\end{center}}

The Euclidean growth rate is followed over a range of 3
magnitudes and breaks down at $B\sim11.1$ mag.  Roughly the same
behavior (but with different limits of completeness) is observed in
different subsamples of the same VS catalog or different VS catalogs.
The magnitude limits of completeness and the number of objects which
meet these criteria are reported in Table 1 for the case of the H
groups.  Although this selection reduces the number of systems
available for the analysis, systematic errors due to incompleteness
are avoided.

\subsection{Results}

In recovering the LF shape and normalization for various group
subsamples, we have used the estimation technique adopted in Paper
II. In particular, the shape has been derived through Turner's (1979)
method and the normalization has been calculated using the relation
$\phi\sbr{s}^*= n\sbr{s} /\int_{-\infty}^{M_c} \phi(M\sbr{s})
dM\sbr{s}$ with $M_c = m\sbr{s}^{lim} - 5\log 500 -15 + 5 \log
h_{75}$, and where $m\sbr{s}^{lim}$ is the limiting apparent magnitude
of completeness of each subsample and $n\sbr{s}$ has been determined
using the minimum variance estimator of Davis \& Huchra (1982).

Our estimate of the VSLF is shown in Figure \ref{figh} for the H and
P1 VS samples.  We have found that a Schechter function with
parameters $\alpha\sbr{s}=-1.4\pm 0.03$, $M\sbr{s}^{*}-5 \log
h_{75}=-23.1 \pm 0.06$ and $\phi\sbr{s}^{*}=4.8 \times
10^{-4}\;\hmpcinvcub$ provides a good fit to data over a broad range
of absolute luminosity ($-24.5 \leq M\sbr{s} - 5 \,log \, h_{75} \leq
-18.5$).  In Figure \ref{figh} we also show the joint distribution of
errors in $\alpha\sbr{s}$ and $M\sbr{s}^{*}$ as derived from the
$\chi^{2}$ matrices of the least squares fit.

\vbox{%
\begin{center}
\leavevmode
\hbox{%
\epsfxsize=8.9cm
\epsffile{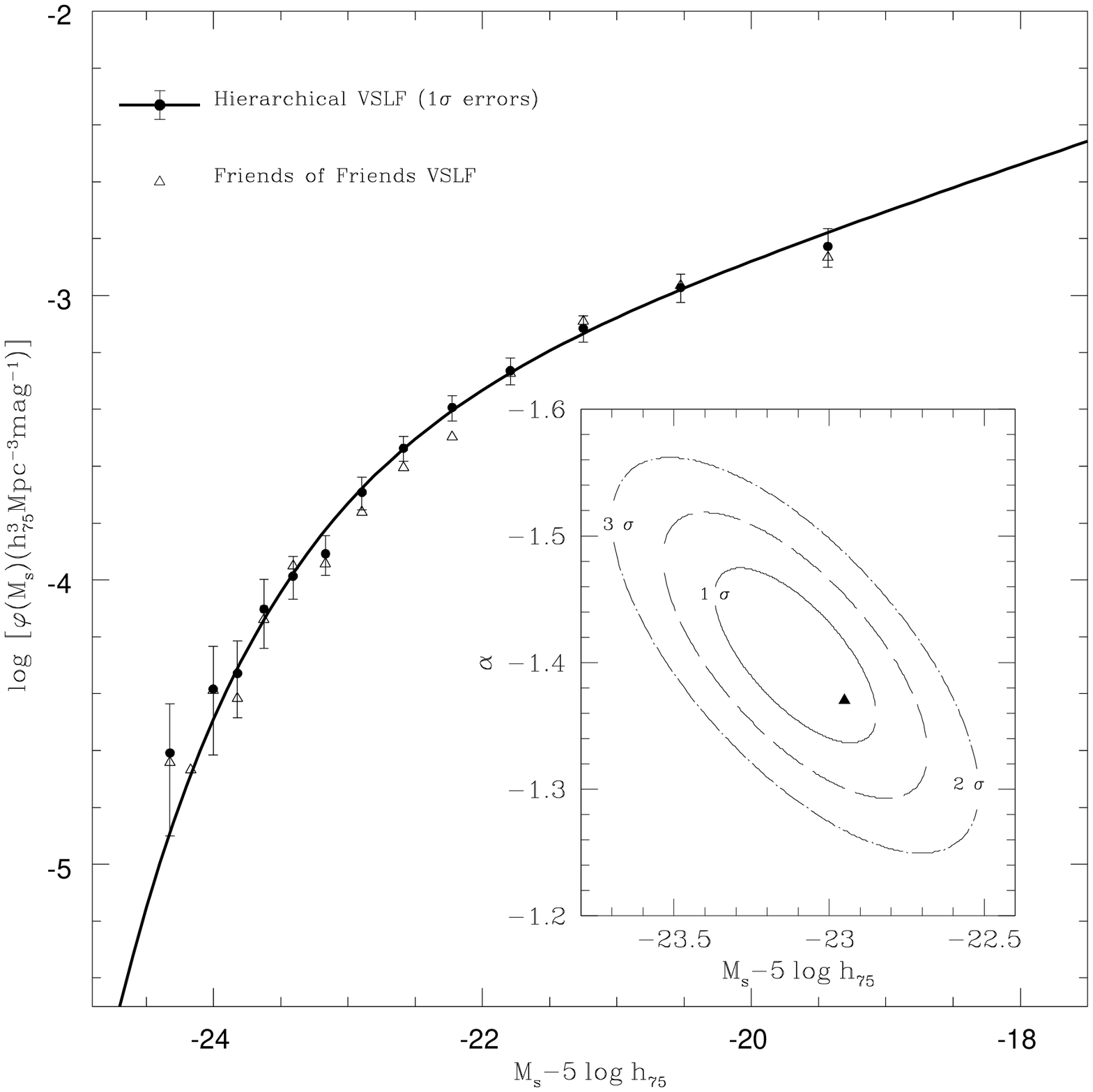}}
\begin{small}
\figcaption{%
The virialized systems luminosity function (differential form) as
derived using NOG hierarchical (H) (solid circles together with $\pm
1\sigma$ error bars) and {\it friends of friends} (P1) (open
triangles) group catalogs. The best fitting Schechter LF function
(solid line) and the 1, 2, 3 $\sigma$ confidence ellipses are also
shown.  The solid triangle represents in the ($\alpha\sbr{s} -
M\sbr{s}^{*}$) space the best-fitting Schechter parameters of the {\it
friends of friends} LF.
\label{figh}}
\end{small}
\end{center}}

Figure \ref{figh} also shows that different catalogs of groups (H and
P1) do not give significantly different LFs.  Moreover, since the P2
groups (obtained by varying the velocity linking parameter) give a LF
which is very similar to the LF obtained from the P1 groups in which
the velocity link parameter is kept fixed, we argue that while other
group properties (e.g.  velocity dispersion) considerably depend on
the values chosen for this parameter (e.g.  Trasarti-Battistoni 1998),
the LF is rather stable and only weakly sensitive to different
choices. Besides we can use redshifts as distance indicators and
obtain a VSLF which is not different from a LF that would be obtained
in a pseudo-real space analysis. For the sake of simplicity, in the
following we shall present results obtained using the H groups.

While the LF is rather insensitive to the group-finding algorithm, it
is somewhat sensitive to the choice of density threshold.  If
thresholds are set much higher than the ones used in this paper, this
yields a larger proportion of field galaxies and, hence, to a steeper
faint-end of the LF, in agreement with the results of MFW.  Marinoni
\& Hudson (2001) show that our adopted density thresholds lead to
systems which are virialized.

In Figure \ref{figi} we show that the VSLF can be economically
described in terms of a smoothly decreasing double-power law of the
following form
\begin{equation}
\phi(L\sbr{s})dL\sbr{s}=\left\{
\begin{array}{r@{\quad{\rm if}\quad}l} \phi\sbr{pl}
\left(\frac{L\sbr{s}}{L\sbr{pl}}\right)^{-1.45          \pm           0.07}
d(\frac{L\sbr{s}}{L\sbr{pl}})   &   L\sbr{s}   <   L\sbr{pl}   \\   \phi\sbr{pl}
\left(\frac{L\sbr{s}}{L\sbr{pl}}\right)^{-2.35                          \pm
0.15}d(\frac{L\sbr{s}}{L\sbr{pl}}) & L\sbr{s} \geq {L\sbr{pl}}
\end{array}
\right.  \end{equation}

\vbox{%
\begin{center}
\leavevmode
\hbox{%
\epsfxsize=8.9cm
\epsffile{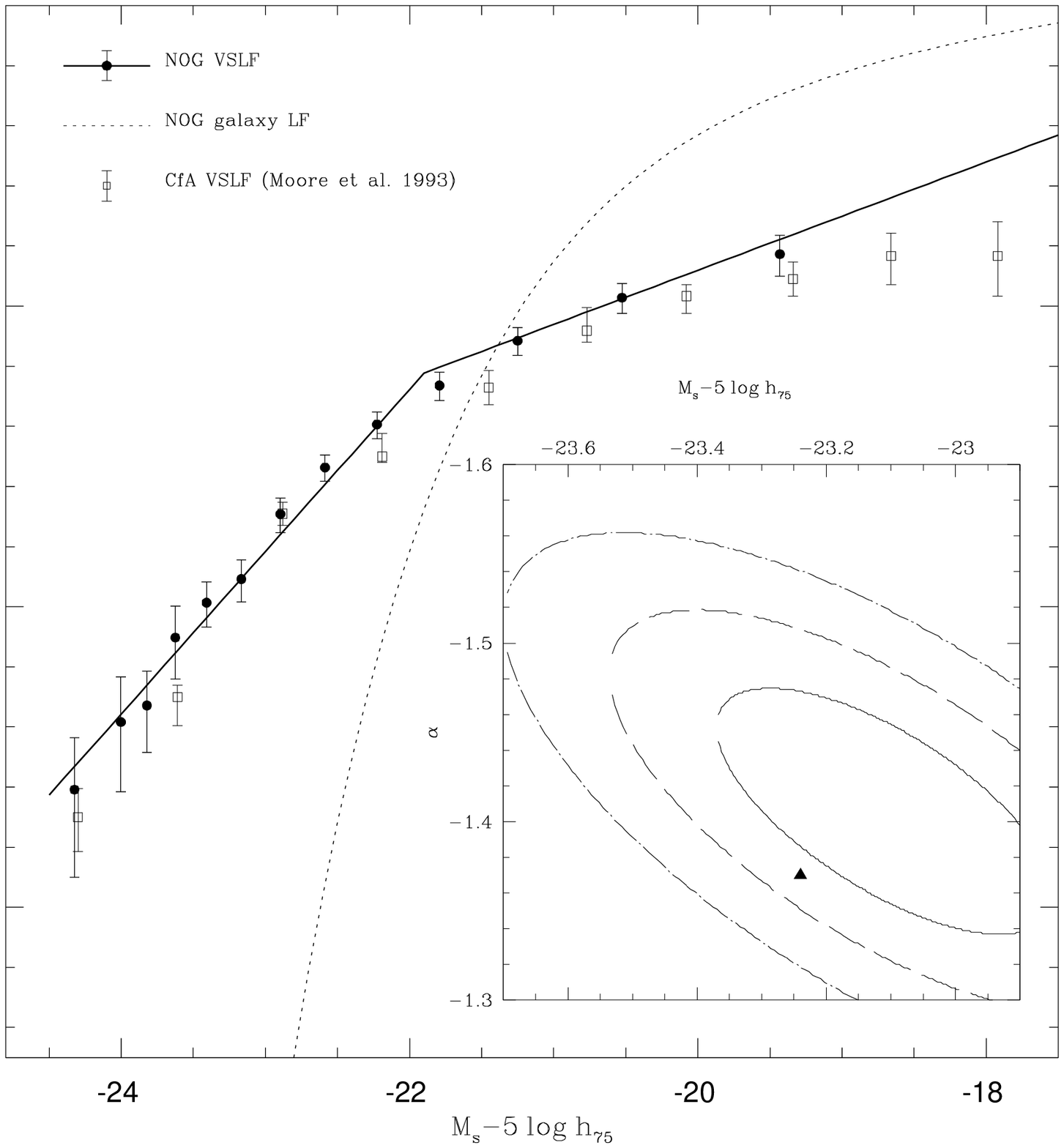}}
\begin{small}
\figcaption{%
The VS luminosity function is plotted (solid circles) and compared to
the CfA VSLF of MFW (squares) and to the NOG galaxy LF (dotted
line). Solid lines represent a double power law fit to data. The 1, 2,
3 $\sigma$ confidence ellipses of the best fitting Schechter LF are
also shown.  The solid triangle represents in the ($\alpha\sbr{s} -
M\sbr{s}^{*}$) space the best-fitting Schechter parameters of the CfA
group LF as given by MFW (after having transformed Zwicky magnitudes
into the B system used in the NOG).
\label{figi}}
\end{small}
\end{center}}

\noindent where $L\sbr{pl}=8.5 \times 10^{10} h_{75}^{-2} L_{\odot}$,
corresponding to $M\sbr{pl} = -21.85$ and $\phi\sbr{pl}=6.5 \times
10^{-4}\;\hmpcinvcub$.  In contrast with the galaxy LF, which has a
sharp break at $L^*\sbr{gal}$, the VSLF shows a more slowly-varying
behavior.  In the limit of low luminosities, the VSLF approaches the
NOG galaxy LF as expected, since these ``systems'' are single field
dwarf galaxies.

The global mean density of the VS distribution is $n \sim 2.2 \times
10^{-2} \hmpcinvcub$ for systems with a magnitude brighter than
$M\sbr{s} -5 \log h_{75} =-15$ and $n \sim 6.2 \times
10^{-3}\hmpcinvcub$ for systems brighter than $M\sbr{s}-5 \log
h_{75}=-18$.

In Figure \ref{figi}, we also show the good agreement ($< 2 \sigma$)
between the CfA groups LF and the NOG VSLF.  In order to compare our
results (which use the corrected $B$ total magnitude in the RC3
system) to the results of MFW (which are based on the Zwicky magnitude
system, $B_z$), we have made the following transformations: $B -
B_{z}=-0.35$ mag (Auman et al.\ 1989) and $B_{(corr)} - B = -0.25$ mag
(for the internal absorption, see Paper II).

The two volumes are largely independent, with NOG being shallower and
wider in area than the CfA1 volume studied by MFW.  The agreement
between the two VSLFs is reassuring because the CfA groups have been
identified using a variant of the P algorithm with different selection
parameters 

\vbox{%
\begin{center}
\leavevmode
\hbox{%
\epsfxsize=8.9cm
\epsffile{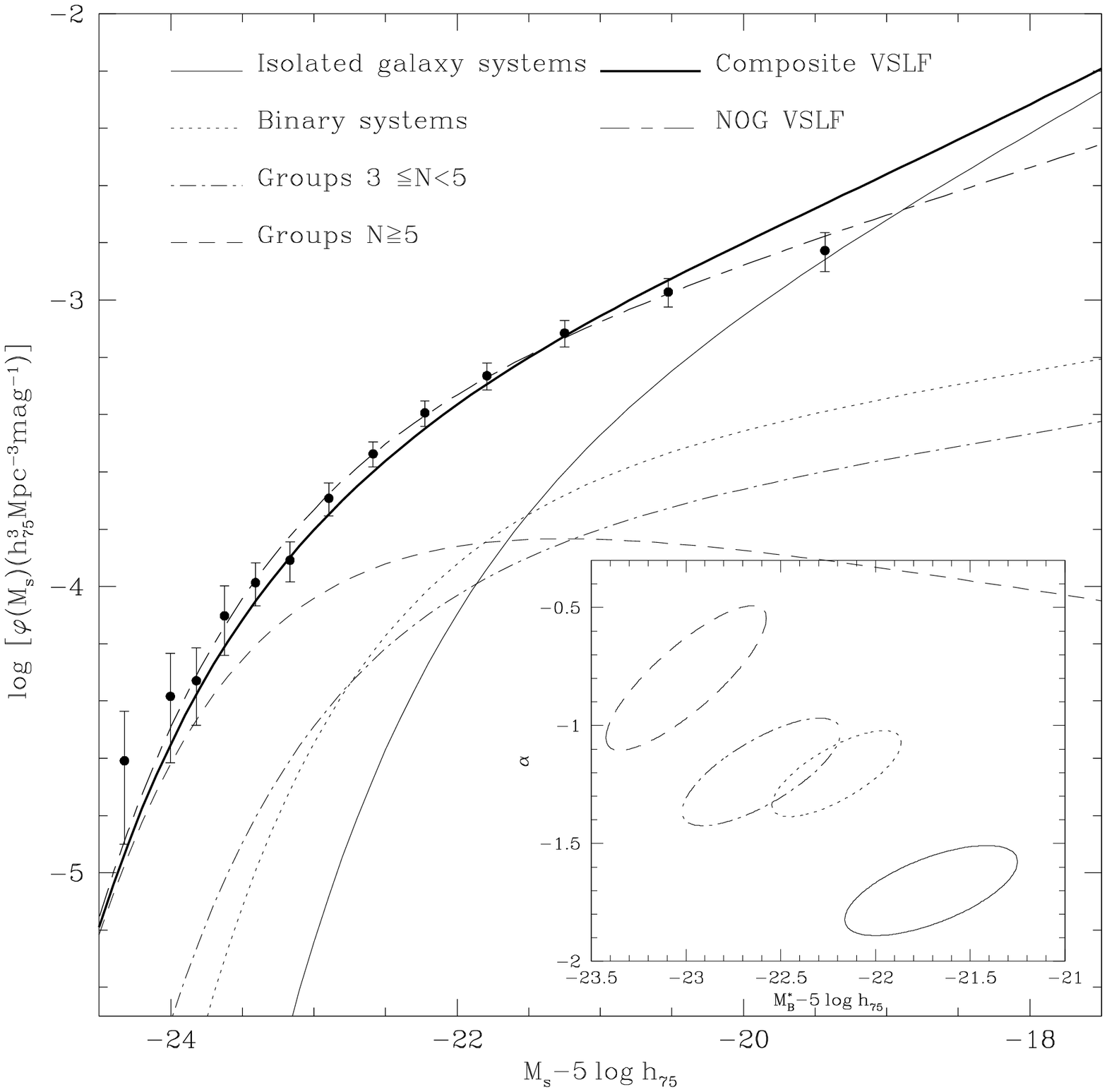}}
\begin{small}
\figcaption{%
Composite group LF obtained by summing the contribution from halos
hosting visible systems of different richness.  Points represent the
VSLF. The corresponding 1 $\sigma$ error ellipses are plotted in the
inset.
\label{figm}}
\end{small}
\end{center}}

and scaling laws, as well as a different luminosity
correction and a different method for estimating the LF.

Moreover, if we consider the GT VSLF, we see that their bright-end
slope has a power-law behavior ($L\sbr{s}^{-7/3}$) with essentially
the same slope of our LF. However, due the lack of small systems with
low surface density enhancement with respect to the background, their
data show a faint end which is flatter than our LF and that of MFW.
Our results are in marginal agreement with the LF derived by Bahcall
(1979) who finds a significantly flatter slope ($\sim 3 \sigma$) at
the faint end.

In Figure \ref{figm}, we have estimated the LF for different range of
group richness from isolated galaxies to groups with N$\geq 5$
members.  The best-fit parameters are given in Table 1.  While field
galaxies dominate the low-luminosity portion of the VSLF, richer
systems contribute in a progressive way to the bright end of the VS
LF.  We also show the total LF defined as the sum over the LFs
pertaining to different samples.

The remarkable flatness of the luminosity-weighted VSLF was first
noted by GT.  A significant contribution to the total luminosity
density of the universe is made by both small and large systems (see
Figure \ref{figo}).  The characteristic luminosity of the NOG {\em
galaxy\/} LF is $L\sbr{gal}^{*}=2.7 \times 10^{10}
h_{75}^{-2}L_{\odot}$ We find that after correcting for the unseen
luminosity, systems with total luminosity $L\sbr{s}<L\sbr{gal}^*$ make
up 25\% of the total luminosity density $\rho_L$ of the universe,
systems with $L\sbr{s}< 5 L\sbr{gal}^*$ and $L\sbr{s}<10 L\sbr{gal}^*$
contribute by 58\% and 75\% respectively, while a substantial fraction
of luminosity density (10\%) is contributed by large systems with more
than $30L\sbr{gal}^*$.

\vbox{%
\begin{center}
\leavevmode
\hbox{%
\epsfxsize=8.9cm
\epsffile{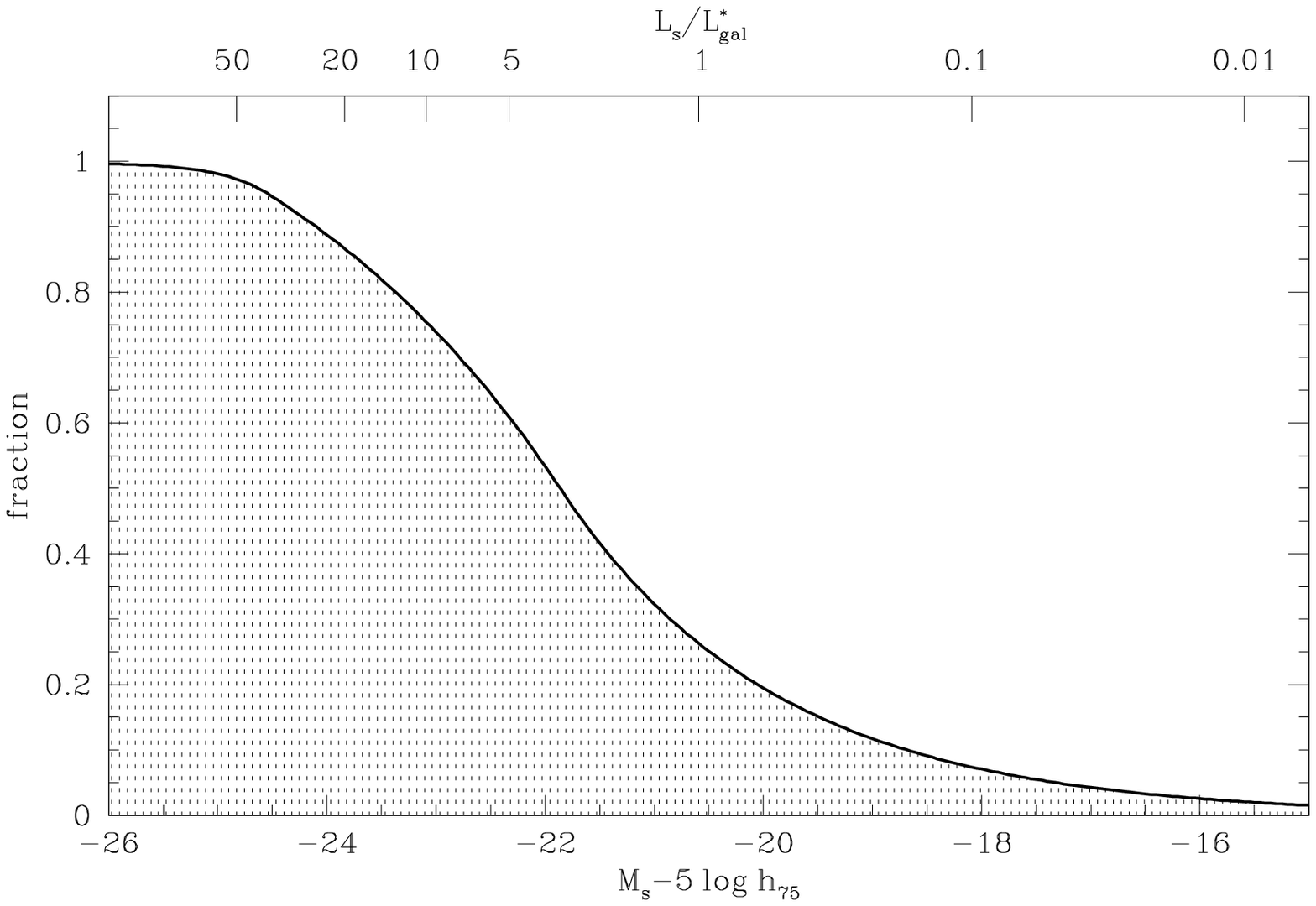}}
\begin{small}
\figcaption{%
Fraction of the total luminosity density from systems with $L <
L\sbr{s}$ as derived from the VSLF.  $L^{*}\sbr{gal}$ is the
characteristic luminosity of the NOG galaxy LF ($M^{*}\sbr{gal}- 5 \log
h_{75}=-20.61$) while $L\sbr{s}$ $(M\sbr{s})$ is the absolute
luminosity of galaxy systems.
\label{figo}}
\end{small}
\end{center}}

\subsection{Environmental Effects on the Luminosity Function of 
Virialized Systems}

The mass function of dark matter halos is expected to show a
dependence on the large-scale environment, as predicted by e.g. Mo \&
White (1996) using an extension of the Press-Schecter formalism, and
confirmed in N-body cold dark matter simulations (see also Lemson \&
Kauffmann 1999). The sense of the predicted trend is that dark matter
halos in overdense regions should be biased toward the high masses
than those forming in lower density regions.  Note that the extended
Press-Schecter theory makes predictions for the clustering of
virialized dark matter halos, which are typically groups or clusters
rather than individual galaxies.

Many workers have looked for such an effect by studying the
luminosities of {\em galaxies\/} as a function of environment, with
conflicting results (e.g., Valotto et al.\ 1997; L\'opez-Cruz et al.\
1997; Bromley et al.\ 1998; Ramella et al.\ 1999; Zabludoff \&
Mulchaey 2000; Christlein 2000; Balogh et al. 2001).  Part of this may
be due to different bandpasses used in different studies, the
environmental dependence being strong in the $K$-band (e.g. Balogh et
al 2001) and weak in the $B$-band.  Marinoni (2001) compared NOG
galaxies in regions of high density contrast to those in low density
contrast, and also compared isolated galaxies to members of
systems. He found that the $B$-band luminosity function of galaxies as
well as the luminosity functions of early and late morphological types
are consistent with no dependence on environment, although a weak
trend (differences in $M_*$ of a few tenths of a magnitude) is also
permitted by the data.

It is clear that at the very bright end, more luminous galaxies are
more strongly clustered (Norberg et al. 2001), and hence one should
expect some dependence of the LF on environment. The trend found by
Norberg et al, while highly significant, is rather weak: the biasing
parameter $b/b_* = 0.85 + 0.15 L/L_*$, where $b_*$ is the biasing
parameter of an $L_*$ galaxy.  

\vbox{%
\begin{center}
\leavevmode
\hbox{%
\epsfxsize=8.9cm
\epsffile{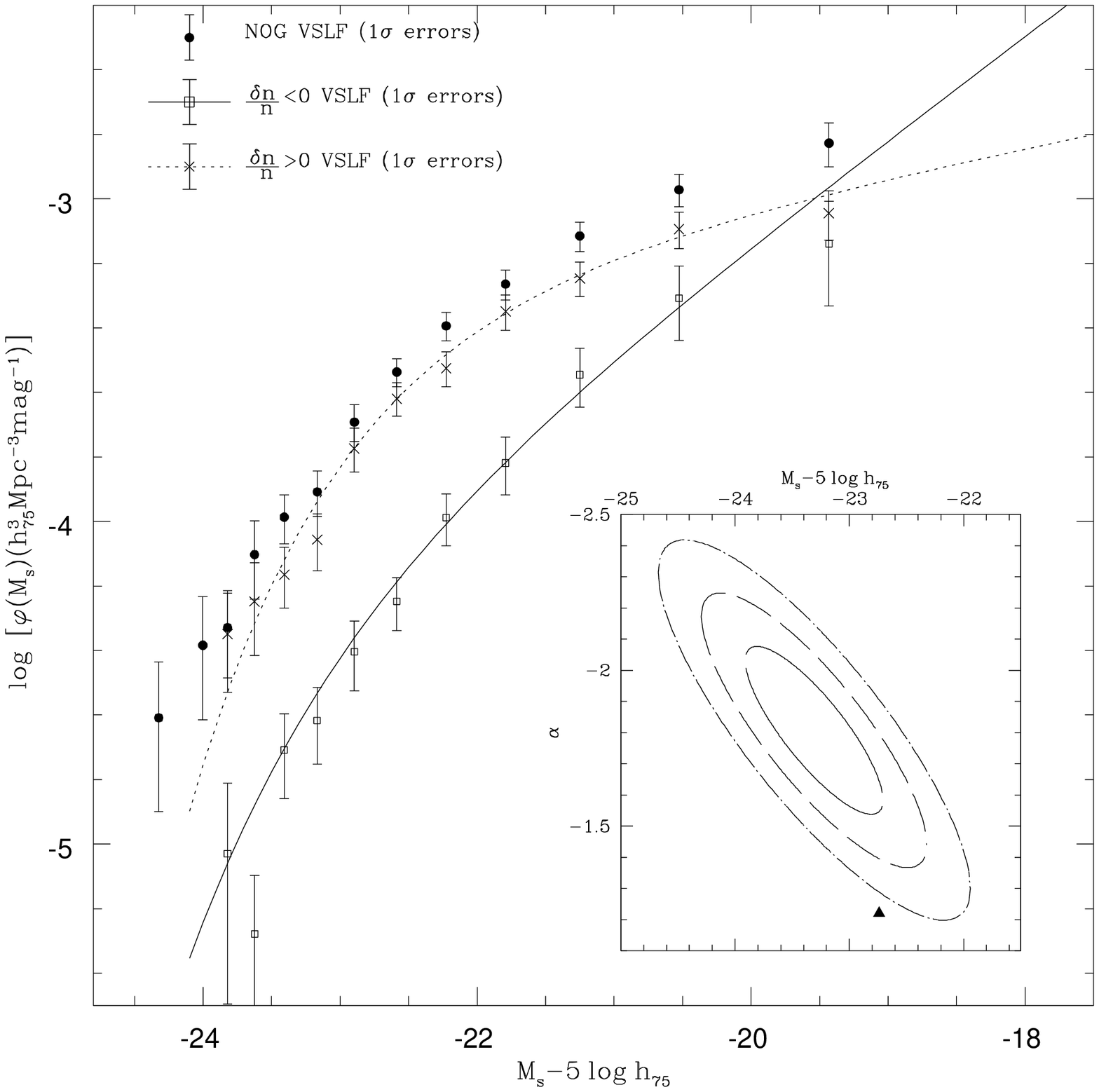}}
\begin{small}
\figcaption{%
The VS luminosity function for hierarchical systems in regions of
different density.  The environmental density parameter $\frac{\delta
n}{n}$ for each halo has been weighted using a Gaussian filter
(eq. \ref{gauss}) with a fixed smoothing length $R\sbr{s}=530$ \kms\
which corresponds to the NOG mean inter-group separation. The inset
shows the 1, 2, 3 $\sigma$ contours for the joint distribution of
errors of the Schechter parameters ($\alpha\sbr{s}, M^{*}\sbr{s}$) of systems in low
density regions. The solid triangle represents in the ($\alpha\sbr{s}
- M\sbr{s}^{*}$) space the best-fitting Schechter parameters of the LF
of systems in high density regions.
\label{figii}}
\end{small}
\end{center}}

Thus at a magnitude brighter than $L*$,
the relative biasing is only 1.22. This leads to a rather weak effect
of when comparing, for example, Schecter function parameters in
overdense and underdense regions.

In contrast to the situation for individual galaxies, we expect that
the total luminosities of virialized systems are closely related to
the dark matter content of the virialized halo.  Systems should
therefore be expected to display the environmental dependence
predicted by the extended Press-Schecter theory. We can test this
prediction as follows.
We assign to each system an environmental density parameter
($\frac{\delta n}{n}$) computed by smoothing the data with a Gaussian
filter (eq. \ref{gauss}) of fixed smoothing length $R\sbr{s}=530$
\kms\ (which corresponds to the mean NOG inter-group separation).
Systems which lie in underdense regions ($\frac{\delta n}{n}<0$; 361
objects) and overdense regions ($\frac{\delta n}{n}>0$; 946 objects)
are differently distributed in luminosity.  Note that the system in
question is excluded from the smoothed density estimate.  

Figure \ref{figii} shows that the hierarchical VSLF is sensitive to the
large-scale environmental density: the luminosity distribution of
halos which inhabit low density regions is biased toward faint
magnitudes.

Both the distributions are well described in term of a Schechter 
function with parameters [-1.8,-23.3] and [-1.2, -22.7] respectively 
in low and high density regions.
The difference in Shechter function parameters is significant at the 
$3 \sigma$ level.                                                    
A signal 
of the same type is also 
present, albeit at a marginal (1.5$\sigma$) significance level 
when we smooth the halo density field on a larger scale
($R\sbr{s}=1000 \kms$).

Thus, in contrast to the situation for galaxies, the VSLF is clearly
dependent on the large-scale environment.  This effect is not the
result of the rather weak dependence of $B$-band galaxy LFs on
environment.  More luminous, and presumably more massive, systems are
more abundant in high-density regions and hence are more strongly
clustered, in agreement with the group correlation function results of
Paper IV.  One astrophysical implication of this result is that the
higher relative biasing of early-type galaxies is then a byproduct of
the environmental dependence of virialized halos coupled with the
morphology-density relation.

\section{Summary and Conclusions}

The traditional approach has been to classify non-linear structures as
being either galaxies, groups or clusters.  In contrast to either
galaxies or rich clusters of galaxies, however, groups have received
little attention in the literature, despite the fact that they these
virialized objects contain most of the luminosity, and hence
presumably 
most of the 
mass, in the universe.  Our approach has been to treat all
virialized ``systems'', from isolated single galaxies to rich clusters
of galaxies, as a continuum,
and to consider the 
total optical luminosity of these systems.

We have analyzed the NOG group catalog (Paper III) in order to correct
the total B-band luminosities of these systems for the luminosity of
galaxies below the magnitude limit of the sample and for the
completeness of subsamples of the group catalog, and have used this to
measure the luminosity function of virialized systems.  We verify that
the luminosity function is insensitive the choice of the group-finding
algorithm.  It is also insensitive to the value of the velocity link
parameter adopted in the percolation algorithm, whereas it is somewhat
sensitive (at the faint end) to the density contrast at which groups
are defined.  Our luminosity function is in good agreement with that
of MFW, who used a quite different approach to its determination and
relied on groups identified in the CfA1 survey.

The luminosity function of virialized systems is well described by a
double-power law, $\phi(L\sbr{s})dL\sbr{s} \propto L\sbr{s}^{-1.45\pm
0.07}dL$ for $L\sbr{s}<L\sbr{s}^*$ and $\phi(L\sbr{s})dL\sbr{s}
\propto L\sbr{s}^{-2.35 \pm 0.15}dL\sbr{s}$ for $L\sbr{s}>L\sbr{s}^*$
with $L\sbr{s}^{*}=8.5 \times 10^{10} h_{75}^{-2} L_{\odot}$, over a
broad range of absolute luminosity ($-24.5 \leq M\sbr{s} - 5 \, \log
\, h_{75} \leq -18.5$).  Our results indicate that 25\%, 50\%, and
75\% of the luminosity of the universe is generated in systems with
$L\sbr{s}< L^{*}\sbr{gal}$, $L\sbr{s}< 2.9 L^{*}\sbr{gal}$ and
$L\sbr{s}< 10 L^{*}\sbr{gal}$, respectively; $10$\% of the overall
luminosity density is supplied by systems with $L\sbr{s}> 30
L^{*}\sbr{gal}$.

Finally, we find that the luminosity function of systems depends on
the large-scale environment, in the sense that halos distributed in
low-density regions host preferentially low-luminosity systems, while
there is an excess of bright virialized systems in high-density
regions.

\acknowledgements

CM would like to acknowledge his Ph.D. adviser GG, to whom this paper
is dedicated, for his constant help and guidance during this project.
He will be missed. 

We also wish to thank C.M.  Baugh, A.J.  Benson, M. Davis,
R. Giovanelli, M.  Haynes, M.  Mezzetti and 
P.  Monaco for interesting conversations.  This work has been
partially supported by the Italian Ministry of University, Scientific
and Technological Research (MURST) and by the Italian Space Agency
(ASI).  MJH acknowledges support from the NSERC of Canada.

\clearpage

\scriptsize{
\begin{deluxetable}{lccccc}
\tablewidth{0pc}
\tablecaption{Luminosity function parameters for various samples 
of systems 
\label{tab1}}
\tablehead{
\colhead{   }  &
\colhead{All systems}  &
\colhead{Isolated} &
\colhead{Binary } &
\colhead{Groups $3 \leq N_g<$5} &
\colhead{Groups $N_g\geq$5} \\
\colhead{   }  &
\colhead{}  &
\colhead{galaxy} & 
\colhead{systems} &
\colhead{ } &
\colhead{ } \\
\colhead{   }  &
\colhead{}  &
\colhead{``systems''} &
\colhead{} &
\colhead{ } &
\colhead{ } \\
}

\startdata
$m\sbr{lim}$            &$ 12.        $&$ 12.       $ &$ 12.         $ &$ 11.5       $	&$ 11.1 $ \\
$N\sbr{s}$            &$ 1307       $&$ 377     $ &$ 450          $ &$ 247        $	&$ 187  $ \\
$V/V\sbr{max}$          &$ 0.47        $&$ 0.61      $ &$ 0.54         $ &$ 0.49       $	&$ 0.31 $ \\
$\alpha$             &$-1.40\pm 0.03 $&$ -1.7 \pm 0.12$ &$ -1.2\pm 0.12  $ &$ -1.2 \pm 0.15 $	&$ -0.8 \pm 0.20$ \\
$M_B^{*} -5 \log h_{75}$ &$-23.1\pm0.06$&$ -21.7 \pm 0.30$ &$ -22.2 \pm 0.22$ &$ -22.6 \pm 0.27$	&$ -23.0 \pm 0.28 $ \\
$\phi^{*}(10^{-4} \hmpcinvcub)$   &$4.8         $&$ 3.9             $ &$2.9              $ &$1.6$&$2.7 $\\
$\rho_{L}(10^8 L_{\odot} \hmpcinvcub) $          &$ 2.0            $&$  0.86           $ &$0.39              $ &$0.32	       $	&$ 0.59 $\\

\enddata
\end{deluxetable}
}

\clearpage
\end{document}